# Bethe Free Energy Approach to LDPC Decoding on Memory Channels

Jaime A. Anguita, *Member, IEEE,* Michael Chertkov, *Member, IEEE,* Mark A. Neifeld, *Member, IEEE,* and Bane Vasic, *Senior Member, IEEE,*

*Abstract*—We address the problem of the joint sequence detection in partial-response (PR) channels and decoding of low-density parity-check (LDPC) codes. We model the PR channel and the LDPC code as a combined inference problem. We present for the first time the derivation of the belief propagation (BP) equations that allow the simultaneous detection and decoding of a LDPC codeword in a PR channel. To accomplish this we follow an approach from statistical mechanics, in which the Bethe free energy is minimized with respect to the beliefs on the nodes of the PR-LDPC graph. The equations obtained are explicit and are optimal for decoding LDPC codes on PR channels with polynomial $h(D) = 1 - \alpha D^n$ ($\alpha$ real, $n$ positive integer) in the sense that they provide the exact inference of the marginal probabilities on the nodes in a graph free of loops. A simple algorithmic solution to the set of BP equations is proposed and evaluated using numerical simulations, yielding bit-error rate performances that surpass those of turbo equalization.

## I. INTRODUCTION

MOST transmission media used in current digital communication systems exhibit a non-uniform frequency response. This non-uniform response, which may manifest in amplitude and/or in phase, introduces distortion to the transmitted signal. This distortion induces a spreading of a symbol waveform beyond its allocated time slot. As a result, a sequence of consecutive symbols transmitted through the channel experiences overlaps of their waveforms, such that the individual symbols are no longer identifiable at the receiver. This symbol overlap is commonly referred to as inter-symbol interference (ISI). Channel ISI is an undesirable feature as it usually increases the complexity of the receiver and may increase the probability of symbol error in the detection process. ISI is observed, for instance, in fiber-optic channels, in the read-back process of magnetic and optical recording channels, and in radio-frequency wireless channels [1]–[3].

In many cases, channel ISI can be represented by a linear filter. In a discrete-time system, this filter gives rise to a state-dependent response, where a channel output is a linear combination of the past transmitted symbols. The number of past symbols affecting the output symbol is denoted as *ISI length* or *memory length*. The memory length is one of the dominant factors determining the ability of a receiver to correctly and efficiently detect symbols in the presence of ISI and receiver noise, and a common way to reduce it is equalizing the channel to a target *partial-response* (PR) channel [4]–[6]. PR channels are designed with short ISI length, to simplify symbol decoding. A discrete-time PR-equalized channel is usually characterized by its PR *polynomial* or *target* $h(D)$, where $D$ denotes the symbol delay. A PR target must be chosen carefully in order not to boost noise to intolerable levels at frequencies for which the channel response is weak. Symbols transmitted over a PR channel are typically decoded using a maximum likelihood (ML) sequence detector [7], like the Viterbi algorithm, or using a maximum a posteriori (MAP) symbol detector [8], like the Bahl, Cocke, Jelinek, and Raviv (BCJR) algorithm. Errors made in the detector may be controlled by an *error correction code*, typically a linear block code.

A class of linear block codes that has received much recognition in the last decade is that of *low-density parity-check* (LDPC) codes. LDPC codes have been shown to give outstanding bit-error rate (BER) performance while featuring a simple encoding and decoding algorithm based on belief propagation [9]–[13]. The belief-propagation algorithm, which operates on a graphical representation of the parity-check matrix of a code, involves passing messages (received bit probabilities or likelihoods) from variable nodes to checks nodes, and vice-versa. This message-passing scheme operates locally on bits and checks and can, consequently, lend itself to low-complexity hardware implementations [10].

Even though knowledge of the channel PR polynomial could be exploited by the error correction decoder, in the state-of-the art receivers symbol detection and error correction are performed separately due to speed and complexity constraints. Roughly, the idea of such sub-optimal algorithms is to provide ISI-free channel symbols as well as their likelihoods to an error-correction decoder. The independence among input symbol likelihoods –assuming that no other a-priori information is available to the decoder– is a necessary condition for successful decoding of LDPC codes by means of the message passing algorithms mentioned above, an example of which is the sum-product algorithm. For a detailed discussion on the sum-product algorithm (or its variants) the reader is referred to [14]. Most of the proposed alternatives for decoding LDPC codes over ISI channels involve the use of the BCJR algorithm

J.A. Anguita was with the Department of Electrical and Computer Engineering, University of Arizona, AZ, USA. He is now with the School of Engineering, Universidad de los Andes, Santiago, Chile. E-mail: janguita@miuandes.cl
M. Chertkov is with the Complex Systems Group, Los Alamos National Laboratory, NM.
M. Neifeld and B. Vasic are with the Department of Electrical and Computer Engineering, University of Arizona, AZ.

(or its variants) followed by the sum-product algorithm. A significantly more efficient approach is to use the output symbol likelihoods of the sum-product algorithm as extrinsic information to improve the performance of the sequence detector in an iterative feedback scheme. This is known as *turbo equalization*, and is currently the most effective known algorithm to decode LDPC codes on PR channels [15]–[18]. Turbo equalization entails, however, high complexity and, because of the sequential nature of the BCJR algorithm, may lead to a significant delay.

Simultaneous channel ISI removal and error-correction decoding is preferred if high bit rates and/or short decoding delays are sought. In this direction, noteworthy contributions have been made by Kurkoski, Siegel, and Wolf [2], and by Pakzad and Anantharam [19]. In the former, a PR channel detector performs parallel symbol message-passing between the PR channel state nodes and the variable nodes for a prescribed number of iterations (graphical models for LDPC codes and ISI channels will be introduced in Section 6.2). The resulting symbol likelihoods are later passed to the LDPC decoder. After a number of iterations of the sum-product algorithm, the LDPC decoder feeds its output likelihoods back to the channel detector, thus completing a turbo iteration. The approach in [2] is, therefore, very similar to turbo equalization, but with the advantage of a significantly shorter delay thanks to the parallel channel detector. Nevertheless, the method can only attain the performance of the BCJR-based turbo equalization algorithm if the number of iterations in the PR channel detector is equal to the LDPC codeword length.

It is well known that the inference of marginal probabilities in a graph can only be uniquely accomplished if the graph is a tree. Since the decoding of LDPC codes corresponds to an inference problem for which the graph has loops, the convergence of the message-passing algorithm is not guaranteed. In this regard, the decoding strategy proposed in [19] seeks to redefine the underlying graphical model –comprised by LDPC nodes/checks and PR nodes– in terms of a set of super-nodes or regions. The belief propagation algorithm is applied to the new, region-based graphical model, thus implementing the Generalized Belief Propagation (GBP) strategy introduced in [20], [21]. It was shown in [19], that the GBP-based approach can outperform the decoding approach of [2]. However, one should also be aware of two important caveats of the GBP-based algorithm. First, the selection of regions is not an unambiguous process, but rather a heuristic strategy, lacking a rigorous justification. Therefore, the quality of the algorithm is not universal, but case- (e.g. code-) specific, and it can only be judged upon simulations. Second, increasing the size of the regions improves GBP but it is also computationally expensive, as the overhead grows exponentially with the region size.

This paper addresses the problem of joint detection and error correction in PR channels. We present, for the first time, a derivation of the belief propagation (BP) equations in a LDPC-coded PR channel. These are obtained by minimizing the Bethe free energy, which is equivalent to performing the exact inference [21] if the graph is loop-free. The derived equations give an explicit solution to the decoding of LDPC codes on general PR channels with pair-wise ISI (i.e., those in which each observed symbol depends on two transmitted symbols), that are corrupted by additive white Gaussian noise. This solution is exact if the LDPC part of the graph does not contain loops and allows a fully parallel implementation on the symbols. The equations reduce to the well-known BP equation for the memoryless channel [14] in the absence of ISI. We also present a simple yet powerful algorithmic solution to the PR-BP equations. The algorithm features a fully parallel implementation, in the sense that channel detection and LDPC decoding are simultaneously performed on each symbol (after a complete codeword has been received). We evaluate the performance of this algorithm on some LDPC codes over the Dicode channel, for which $h(D) = 1 - D$, and find that it outperforms the turbo equalization algorithm. We illustrate the smooth convergence of PR-BP algorithm towards the ISI-free channel case by evaluating its performance over a channel with $h(D) = 1 + 0.5D$.

The remainder of the paper is organized as follows. In Section II we briefly introduce LDPC codes and their graphical representation. We also introduce a graphical model for LDPC codes on a linear ISI channel. In Section III the derivation of the BP equations from the Bethe free energy is given, and an iterative decoding algorithm to solve the equations is proposed. In Section IV we present a numerical evaluation of the bit-error rate (BER) versus signal-to-noise ratio (SNR) of several LDPC codes over the Dicode channel. A comparison against the BER performance of a turbo equalizer is also offered to display the excellent convergence of the PR-BP algorithm, particularly with medium- to low-rate codes. A brief comparison of complexity between our approach and that of turbo equalization is offered at the end of Section IV. Finally, Section V summarizes our results.

## II. PRELIMINARIES ON GRAPHICAL MODELS

### A. Graphical representation of a LDPC code

Let $\boldsymbol{x} = \{x_1, x_2, \ldots, x_N\}$ denote an ordered set of variables each of which can take values from a finite alphabet $\mathcal{B}$. Let $g$ indicate a function of these variables. A *configuration* of $\boldsymbol{x}$ denotes a particular realization of $\boldsymbol{x}$ from the domain $\mathcal{S} = \mathcal{B}^N$, referred to as the *configuration space*. A *marginal* function $g_i(x_i)$ is a function such that for each $\gamma \in \mathcal{B}$, $g_i(\gamma)$ is found by summing $g(\boldsymbol{x})$ over all those configurations for which $x_i = \gamma$. Namely, the marginal function $g_i(x_i)$ is expressed by [14]

$$g_i(x_i) = \sum_{\boldsymbol{x} \setminus x_i} g(x_1, x_2, \ldots, x_n)$$

where $\boldsymbol{x} \setminus x_i$ denotes that the summation is over all variables in $\boldsymbol{x}$ except $x_i$. Let us assume that $g(\boldsymbol{x})$ can be expressed as a product of functions $f_\alpha$ whose arguments $\boldsymbol{x}_\alpha$ are subsets of $\boldsymbol{x}$, and $\alpha$ is an element of the index set $A$. We write $g(\boldsymbol{x})$ as

$$g(x_1, x_2, \ldots, x_n) = \prod_{\alpha \in A} f_\alpha(\boldsymbol{x}_\alpha), \quad (1)$$

and the marginal $g_i(x_i)$ can be written as

$$g_i(x_i) = \sum_{\boldsymbol{x} \setminus x_i} \prod_{\alpha \in A} f_\alpha(\boldsymbol{x}_\alpha). \quad (2)$$

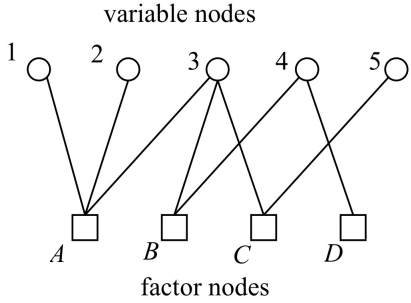

Fig. 1. Factor graph or bipartite graph. Circles correspond to the variable (bit) nodes and the squares correspond to the factor (check) nodes.

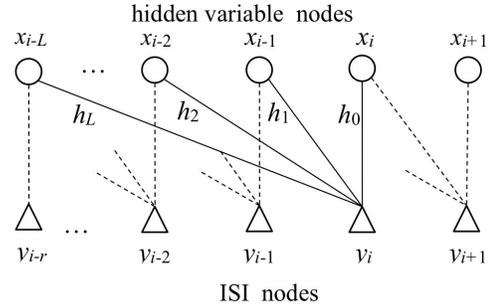

Fig. 2. Graphical representation of a linear ISI channel. The output nodes $y_i$ are a linear combination of the uncorrelated nodes $x_i$ plus additive receiver noise.

A factor graph is a bipartite graph whose configuration is determined via (1) [22], [23]. In a factor graph, variables xi are symbolized by variable nodes; factor functions $f_\alpha$ are symbolized by factor nodes; and the dependence of a function on a variable is symbolized by an edge joining the two. It is not difficult to see that every factor graph is a tree. Figure 1 depicts an example of a factor graph, in which the variable nodes are represented by circles and the factor nodes are represented by squares. This graph has five variable nodes and four factor nodes. Its structure corresponds to the functional expression

$$g(x_1, x_2, x_3, x_4, x_5) = f_A(x_1, x_2, x_3) f_B(x_3, x_4)$$
$$\times f_C(x_3, x_5) f_D(x_4). \quad (3)$$

A LDPC code is a linear block code specified by its parity-check matrix $\mathbf{H}$. This matrix has elements from the set $\{0,1\}$ and is sparse, i.e., the number of elements 1 is much smaller than the number of elements 0. $\mathbf{H}$ is said to be regular if it features uniform column and row weight; otherwise, it is called irregular. A parity-check matrix $\mathbf{H}$ of $M$ rows and $N$ columns and rank $\text{rank}(H)$ defines a code $C$ with block length $N$ and rate $(N - \text{rank}(H))/N$. Each row of $\mathbf{H}$ defines a parity check equation. A 1 in row $j$ and column $i$ indicates that variable $x_i$ is an argument of the $j$th parity check equation. A codeword of the code $C$ is a configuration of the ordered set of variables $\mathbf{x}$ for which all the parity check equations are satisfied. If the alphabet $\mathcal{B}$ is binary with elements from GF(2), then a codeword of $C$ (in vector representation) satisfies $\mathbf{H}\mathbf{x}^T = 0$ over GF(2), where $\mathbf{0}$ is an all-zero vector. A bipartite graph in which the parity check equations are represented by factor nodes and the variables $x_i$ by variable nodes is referred to as a *Tanner graph*. A value 1 at row $j$ and column $i$ of $\mathbf{H}$ is represented by an edge between variable node i and factor node $j$. We define $q_i$ as the node degree (i.e., the number of connected edges) of variable node $i$ and $p_j$ as the node degree of factor node $j$ [10]. A regular LDPC codes has $q_i = q$, $p_j = p$, $\forall i,j$.

### B. The discrete ISI channel

Consider the transmission of a sequence of symbols $x_i$ in discrete time intervals indexed by $i$. A linear discrete ISI channel relates the output signal $y_i$ with the transmitted signal $x_i$ as

$$y_i = \sum_{j=0}^{L} h_j x_{i-j} + \xi_i \quad (4)$$

where $L$ is the ISI length, $h_0, h_1, \ldots, h_r$ are real-valued channel coefficients, and $\xi_i$ is an additive discrete noise process, which we assume to be white and Gaussian with zero mean and variance $\sigma^2$. The relation in (4) is commonly referred to as the PR channel. The PR channel is usually represented by the polynomial expression

$$h(D) = \sum_{j=0}^{L} h_j D^j \quad (5)$$

where $D$ is the delay operator, such that $D^j x_i = x_{i-j}$. We assume that the polynomial $h(D)$ is normalized so that $h_0 = 1$. Figure 2 shows a graphical representation of (5). Each variable $x_i$ is represented by a *hidden* variable node, that is, a variable node that cannot be observed. An output variable $y_i$ is designated by a triangle in the graph, which we denote as an ISI node. ISI nodes include the contribution of additive noise, but to simplify the graph we choose to leave this contribution implicit. We denote the SNR by $s^2$ and, following [24], is defined as

$$s^2 = \frac{\sum_{j=0}^{L} h_j^2}{\sigma^2}. \quad (6)$$

This definition of SNR accounts for the energy contribution induced by the ISI, so that the energy per symbol is maintained, regardless of the channel coefficients and the memory length $L$.

### C. Graphical representation of a LDPC code on a discrete ISI channel

An LDPC code operating on an ISI channel can have a graphical representation that combines the graphs in Figs. 1 and 2. An example of this is the tripartite graph in Fig. 3, which correspond to a linear code on a discrete ISI channel of length $L$. Provided that the factor nodes in Fig. 3 correspond to the parity check equations of the LDPC code, we may interchangeably refer to them as *check nodes*. We have previously denoted the parity check equations by $f_\alpha$. In addition, we denote the functional representation of the ISI node by $f_\aleph$ ($\aleph$ is the first letter of the hebrew alphabet). The tripartite graph in Fig. 3 serves as the basis for the derivation of the BP equations over a discrete ISI channel. Note that the left-most ISI node in the graph is assumed to be the first output signal observed. The next section presents the derivation of the belief propagation equations.

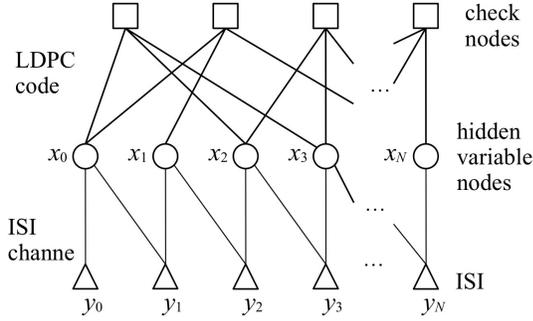

Fig. 3. Factor graph of a LDPC code on a ISI channel with length $L = 1$. Squares, circles, and triangles represent factor nodes, variable nodes, and ISI nodes, respectively.

## III. BETHE FREE ENERGY AND BELIEF PROPAGATION EQUATIONS

We begin this Section by briefly introducing the concept of beliefs and the Bethe free energy. The reader may find a thorough description of the BP algorithm and its relation to the Bethe free energy in [21]. A belief $b_i(x_i)$ at the variable node $i$ is an approximation to the exact marginal function $g_i(x_i)$ [21]. We can extend this definition to the other types of nodes in the graph shown in Fig. 3. The joint belief $b_\alpha(\boldsymbol{x}_\alpha)$ at the set of variables $\boldsymbol{x}_a$ (which in turn corresponds to the belief of the factor node $f_\alpha$) is an approximation to the exact marginal function $g_\alpha(\boldsymbol{x}_\alpha)$. Similarly, the joint belief $b_\aleph(\boldsymbol{x}_\aleph)$ of the set of variables $\boldsymbol{x}_\aleph$, corresponding to the variable nodes connected to the ISI node $\aleph$, is an approximation to the exact marginal function $g_\aleph(\boldsymbol{x}_\aleph)$.

It is of interest to compute the marginal function mentioned above, because they represent the probabilities of the transmitted symbols. However, even with the knowledge of the global function $g(\boldsymbol{x})$, this may be a very difficult computational task. We may use the BP equations to approximate the marginal functions by means of the beliefs.

### A. Bethe free energy approach to the decoding of LDPC codes in a PR channel

The BP equations correspond to the stationary points of a function of the beliefs called the Bethe free energy [21]. The Bethe free energy, expressed as a function of the beliefs $b_\alpha(\boldsymbol{x}_\alpha)$ and $b_\aleph(\boldsymbol{x}_\aleph)$ on the check, and the ISI nodes, respectively, is

$$F[b_\alpha(\boldsymbol{x}_\alpha), b_\aleph(\boldsymbol{x}_\aleph)] = U[b_\alpha(\boldsymbol{x}_\alpha), b_\aleph(\boldsymbol{x}_\aleph)] \\ - H[b_\alpha(\boldsymbol{x}_\alpha), b_\aleph(\boldsymbol{x}_\aleph)], \quad (7)$$

where the Bethe self energy $U$ is

$$U[b_\alpha(\boldsymbol{x}_\alpha), b_\aleph(\boldsymbol{x}_\aleph)] = -\sum_\alpha \sum_{\boldsymbol{x}_\alpha} b_\alpha(\boldsymbol{x}_\alpha) \ln f_\alpha(\boldsymbol{x}_\alpha) \\ - \sum_\aleph \sum_{\boldsymbol{x}_\aleph} b_\aleph(\boldsymbol{x}_\aleph) \ln f_\aleph(\boldsymbol{x}_\aleph) \quad (8)$$

and the Bethe entropy $H$ is

$$H[b_i(x_i), b_\alpha(\boldsymbol{x}_\alpha), b_\aleph(\boldsymbol{x}_\aleph)] = \sum_\alpha \sum_{\boldsymbol{x}_\alpha} b_\alpha(\boldsymbol{x}_\alpha) \ln b_\alpha(\boldsymbol{x}_\alpha) \\ - \sum_\aleph \sum_{\boldsymbol{x}_\aleph} b_\aleph(\boldsymbol{x}_\aleph) \ln b_\aleph(\boldsymbol{x}_\aleph) \\ + \sum_i (q_i - 1) \sum_{x_i} b_i(x_i) \ln b_i(x_i). \quad (9)$$

In a tree-like graph, the Bethe free energy is a concave function on the beliefs such that at its minimum points, $b_\alpha(\boldsymbol{x}_\alpha) = g_\alpha(\boldsymbol{x}_\alpha)$ and $b_\aleph(\boldsymbol{x}_\aleph) = g_\aleph(\boldsymbol{x}_\aleph)$, the desired marginal functions, and $F = F_{\text{free}}$, the free energy. Since it is of interest to interpret the beliefs as probability mass functions, the normalization constraints

$$\sum_{x_i} b_i(x_i) = \sum_{\boldsymbol{x}_\alpha} b_\alpha(\boldsymbol{x}_\alpha) = \sum_{\boldsymbol{x}_\aleph} b_\aleph(\boldsymbol{x}_\aleph) = 1, \quad (10)$$

and the consistency constraints

$$b_i(x_i) = \sum_{\boldsymbol{x}_\alpha \setminus x_i} b_\alpha(\boldsymbol{x}_\alpha) = \sum_{\boldsymbol{x}_\aleph \setminus x_i} b_\aleph(\boldsymbol{x}_\aleph), \quad (11)$$

must be satisfied.

The probability of observing the channel output vector $y$, given the binary input vector $\boldsymbol{x} \in \{-1, 1\}^N$ and the SNR $s^2$, is

$$P(\boldsymbol{y}|\boldsymbol{x}) \propto \exp\left[-\frac{s^2}{2} \sum_{i=1}^N \left(y_i - \sum_{j=0}^L h_j x_j\right)^2\right] \quad (12)$$

provided that the discrete additive noise process is white and Gaussian. After expanding (12) and discarding the constant terms, we obtain

$$P(\boldsymbol{y}|\boldsymbol{x}) \propto \exp\left(\sum_{i=1}^N u_i \, x_i\right) \exp\left(-\sum_{(i,j)}^{1 \leq |i-j| \leq L} Q_{j-i} \, x_j \, x_i\right), \quad (13)$$

where the summation on the right-most exponent is over all pairs $(i, j)$ of distinct bits separated by a distance $(i, j), 1 \leq |i - j| \leq L$, and we have defined

$$u_i = s^2 \sum_{j=0}^L h_j \, y_j, \quad Q_p = s^2 \sum_{k=0}^{|p-L|} h_k \, h_{k+p}. \quad (14)$$

With $u_i$ we symbolize the likelihood of the variable node $i$. For $L = 0$, $u_i$ takes the form of the likelihood of a Gaussian memoryless channel. $Q_p$ accounts for the pair-wise memory.

Following the approach in [21], [25], we would like to write the joint probability distribution function of the random vector $\mathbf{X}$ representing the binary symbols of a codeword with a product of functions such that the probability of the configuration $\boldsymbol{x} = \{x_1, x_2, \ldots, x_N\}$, $p(\boldsymbol{x})$, is given by

$$p(\boldsymbol{x}) = \frac{1}{Z} \prod_{\alpha \in A} f_\alpha(\boldsymbol{x}_\alpha) \prod_{\aleph \in \beth} f_\aleph(\boldsymbol{x}_\aleph) \quad (15)$$

where $\{f_\alpha(\boldsymbol{x}_\alpha)\}$ is a set of $M$ non-negative functions as defined in Section 6.2. Similarly, $\{f_\aleph(\boldsymbol{x}_\aleph)\}$ is a set $\beth$ of $N$ non-negative functions indexed by $\aleph$ whose arguments $\boldsymbol{x}_\aleph$ are subsets of $\boldsymbol{x}$. In (15), $Z$ is a normalization constant given by

$$Z = \sum_{\boldsymbol{x}} \prod_{\alpha \in A} f_\alpha(\boldsymbol{x}_\alpha) \prod_{\aleph \in \beth} f_\aleph(\boldsymbol{x}_\aleph) \quad (16)$$

such that $\sum_{\boldsymbol{x}} p(\boldsymbol{x}) = 1$, i.e., $p(\boldsymbol{x})$ is a probability mass function. The purpose of expressing the joint probability distribution in this fashion is to conveniently represent the factor graph in Fig. 3, in which $\{f_\alpha(\boldsymbol{x}_\alpha)\}$ describes the check nodes of the LDPC parity-check matrix, and $\{f_\aleph(\boldsymbol{x}_\aleph)\}$ describes the observed ISI nodes. By writing

$$f_\alpha(\boldsymbol{x}_\alpha) \equiv \delta\left(\prod_{x_j \in \boldsymbol{x}_\alpha} x_j, 1\right) \exp\left(q_i^{-1} \sum_{x_i \in \boldsymbol{x}_\alpha} x_i\, u_i\right) \quad (17)$$

$$f_\aleph(\boldsymbol{x}_\aleph) \equiv \exp\left(-Q_{|j-i|}\, x_i x_j\right), \quad (18)$$

where $\delta(v, 1) = 1$ for $v = 1$ (the parity-check equation is satisfied) and 0 otherwise, we obtain

$$p(\boldsymbol{x}) = \frac{1}{Z} \prod_{\alpha=1}^{M} \delta\left(\prod_{x_j \in \boldsymbol{x}_\alpha} x_j, 1\right) \prod_{i=1}^{N} \exp(u_i x_i)$$
$$\times \prod_{\substack{(i,j) \\ 1 \leq |i-j| \leq L}} \exp\left(-Q_{j-i}\, x_j\, x_i\right). \quad (19)$$

The Bethe free energy is then minimized with respect to the beliefs $b_i$, $b_\alpha$, and $b_\aleph$ subject to the normalization and consistency constraints in (10) and (11). Namely, we minimize the Lagrangian function

$$\mathcal{L} = U - H + \sum_\alpha \gamma_\alpha \left(\sum_{\boldsymbol{x}_\alpha} b_\alpha(\boldsymbol{x}_\alpha) - 1\right)$$
$$+ \sum_\aleph \gamma_\aleph \left(\sum_{\boldsymbol{x}_\aleph} b_\aleph(\boldsymbol{x}_\aleph) - 1\right)$$
$$+ \sum_i \gamma_i \left(\sum_{x_i} b_i(x_i) - 1\right)$$
$$+ \sum_i \sum_{\alpha \ni i} \sum_{x_i} \lambda_{i\alpha}(x_i) \left[b_i(x_i) - \sum_{\boldsymbol{x}_\alpha \setminus x_i} b_\alpha(\boldsymbol{x}_\alpha)\right]$$
$$+ \sum_i \sum_{\aleph \ni i} \sum_{x_i} \lambda_{i\aleph}(x_i) \left[b_i(x_i) - \sum_{\boldsymbol{x}_\aleph \setminus x_i} b_\aleph(\boldsymbol{x}_\aleph)\right], \quad (20)$$

where $\alpha \ni i$ and $\aleph \ni i$ indicate all indices of the checks and of the ISI nodes connected to bit $i$, respectively, and $\gamma_i, \gamma_\alpha, \gamma_\aleph, \lambda_{i\alpha}(x_i), \lambda_{i\aleph}(x_i)$ are Lagrange coefficients that multiply the normalization constraints (10) and the consistency constraints (11). The minimization of (20) with respect to the beliefs leads to

$$b_\alpha(\boldsymbol{x}_\alpha) = f_\alpha(\boldsymbol{x}_\alpha) \exp\left[-\gamma_\alpha - 1 + \sum_{i \in \alpha} \lambda_{i\alpha}(\sigma_i)\right], \quad (21)$$

$$b_\aleph(\boldsymbol{x}_\aleph) = f_\aleph(\boldsymbol{x}_\aleph) \exp\left[-\gamma_\aleph - 1 + \sum_{i \in \aleph} \lambda_{i\aleph}(x_i)\right], \quad (22)$$

$$b_i(x_i) = \exp\left[\frac{1}{q_i + L - 1}\left(\gamma_i + \sum_{\alpha \ni i} \lambda_{i\alpha}(x_i) + \sum_{\aleph \ni i} \lambda_{i\aleph}(x_i)\right) - 1\right]. \quad (23)$$

Equations (21)-(23) complemented by the normalization and the consistency constraints form a close system of BP equations for the $\lambda_{i\alpha}(x_i)$ and $\lambda_{i\aleph}(x_i)$ coefficients. Following the traditional notation of BP equations in terms of the fields $\eta$ defined on the edges of the factor graph, we have the relations

$$\eta_{i\alpha} \equiv \frac{\lambda_{i\alpha}(+1) - \lambda_{i\alpha}(-1)}{2} + \frac{h_i}{q_i},$$
$$\eta_{i\aleph} \equiv \frac{\lambda_{i\aleph}(+1) - \lambda_{i\aleph}(-1)}{2}, \quad (24)$$

where $\eta_{i\alpha}$ indicates the field going from variable node $i$ to factor node $\alpha$, and $\eta_{i\aleph}$ indicates the field going from variable node $i$ to ISI node $\aleph$. Substituting (24) in (21)-(23) yields the expressions

$$\sum_{\boldsymbol{x}_\alpha} x_i b_\alpha(\boldsymbol{x}_\alpha) = \tanh\left[\eta_{i\alpha} + \tanh^{-1}\left(\prod_{\substack{j \in \alpha \\ j \neq i}} \tanh \eta_{j\alpha}\right)\right], \quad (25)$$

$$\sum_{\boldsymbol{x}_\aleph} x_i b_\aleph(\boldsymbol{x}_\aleph) = \tanh\left(\eta_{i\aleph} - \tanh^{-1}\left(\tanh \eta_{i\aleph} \tanh Q_\aleph\right)\right),$$
$$\text{such that } (i, j) \in \aleph, \quad (26)$$

$$\sum_{x_i} x_i b_i(x_i) = \tanh\left[\frac{1}{q_i + L - 1}\left(\sum_{\alpha \ni i} \eta_{i\alpha} + \sum_{\aleph \ni i} \eta_{i\aleph}\right)\right] \quad (27)$$

By equating the right-hand sides of (25), (26), and (27), and using (24), after some algebraic manipulations we find the BP equations for the ISI channel:

$$\eta_{i\alpha} = u_i + \sum_{\substack{\beta \ni i \\ \beta \neq \alpha}} \tanh^{-1}\left(\prod_{\substack{j \in \beta \\ j \neq i}} \tanh \eta_{j\beta}\right)$$
$$- \sum_{\aleph \ni i} \tanh^{-1}\left(\tanh \eta_{i\aleph} \tanh Q_\aleph\right), \quad (28)$$

$$\eta_{i\aleph} = u_i - \sum_{\substack{\beth \ni i \\ \beth \neq \aleph}} \tanh^{-1}\left(\tanh \eta_{i\beth} \tanh Q_\beth\right)$$
$$+ \sum_{\alpha \ni i} \tanh^{-1}\left(\prod_{\substack{j \in \alpha \\ j \neq i}} \tanh \eta_{j\alpha}\right). \quad (29)$$

where the hebrew letter $\beth$ is used to indicate those ISI nodes that are connected to the variable node $i$ but are not $\aleph$.

It can be observed in (28)-(29) that in the absence of ISI, $Q_\aleph = 0$ and the equations reduce to the well known BP equations for memoryless channels.

Equations (28)-(29) form the exact BP solution to the PR channel with pair-wise ISI, regardless of the distance between the two variable nodes joint by an ISI node. In a factor graph whose PR nodes do not form loops (in the absence of the check nodes), the solution is exact in the sense that the fields determined correspond to the stationary points of the Bethe free energy. Of course, the check nodes of the LDPC code add loops to the graph, and therefore, the convergence of the fields is not guaranteed. The BP equations above can be used to decode the message symbols of a LDPC code over a discrete PR channel with polynomial

$$h(D) = 1 - \alpha D^n, \tag{30}$$

where $-1 \leq \alpha \leq 1$ and $n$ is a non-negative integer. For ISI involving more than two variable nodes, the solution is suboptimal even in the absence of the check nodes because the ISI nodes and the variable nodes will form loops.

The BP equations are nonlinear and the roots can be evaluated with the method of preference. In the next subsection we describe a simple iterative procedure to solve the BP equations, analogous to the message-passing algorithm applied to memoryless channels.

### B. Algorithmic solution to the BP equations

The nonlinear PR-BP equations described above can be solved using any nonlinear minimization algorithm. However, in the interest of finding a simple and fully-parallel decoding algorithm, we follow an iterative approach similar to that of message-passing for LDPC codes on memoryless channels. Equations (28)-(29) can be iteratively decoded using the following algorithm:

$$\eta_{i\alpha}^{(n+1)} = u_i + \sum_{\substack{\beta \ni i \\ \beta \neq \alpha}} \mu_{i\beta} - \sum_{\aleph \ni i} \zeta_{i\aleph} \tag{31}$$

$$\eta_{i\aleph}^{(n+1)} = u_i - \sum_{\substack{\beth \ni i \\ \beth \neq \aleph}} \zeta_{i\beth} + \sum_{\alpha \ni i} \mu_{i\beta} \tag{32}$$

with

$$\mu_{i\beta} = \tanh^{-1} \left( \prod_{\substack{j \in \beta \\ j \neq i}} \tanh \eta_{j\beta}^{(n)} \right) \tag{33}$$

$$\zeta_{i\aleph} = \tanh^{-1} \left( \tanh \eta_{i\aleph}^{(n)} \tanh Q_\aleph \right) \tag{34}$$

The superscript $(n+1)$ refers to the value of the field $\eta$ at iteration step $n+1$. Decoding starts when all ISI symbols associated with a LDPC codeword have been received. The algorithm is initialized by setting all $\mu_{i\beta}, \zeta_{i\aleph}$ to zero. It is assumed that the symbols transmitted prior and posterior to the codeword are on a known state. Using terminating nodes with known states is not strictly necessary when the channel has memory length $L = 1$, as convergence is enforced by the check equations. However it helps achieving a faster convergence and it is also common practice in the evaluation of decoders on PR channels. The fields in (31) and (32) are evaluated using the values $\mu_{i\beta}, \zeta_{i\aleph}$, computed in the previous iteration to replicate a fully-parallel architecture at each iteration. After every iteration, the likelihood $\Lambda_i$ on each variable node $i$ is computed as

$$\Lambda_i = u_i + \sum_{\alpha \ni i} \mu_{i\alpha} \tag{35}$$

and the codeword is checked. Note that the summation is in this case over all the check nodes a connected to variable node $i$. In the next section we present a BER evaluation of some LDPC codes using this iterative algorithm on a Dicode channel.

## IV. NUMERICAL SIMULATIONS

As we have already mentioned, the exact solution given above does not guarantee convergence to a valid codeword because linear block codes generate graphs with loops. It is well-known, however, that very good convergence is generally observed using the sum-product algorithm on graphs with loops such as those generated by LDPC codes [12]. We expect this to be also true for the BP equations presented here. To illustrate this, we have numerically performed the transmission and decoding of random LDPC codewords on a Dicode channel, given by the polynomial $h(D) = 1 - D$ by means of Monte Carlo simulations. We have done this for various LDPC codes using the algorithm in (31)-(34) with the initialization given in (14). Clearly, for the Dicode channel $h_0 = 1$, $h_1 = -1$, and $Q_1 = -s^2$. Each channel observation $y_i$ is computed using (4), in which $x_i$ is a LDPC-coded binary symbol that takes values from $\{-1, +1\}$, and $\xi_i$ is a AWGN sample from the normal distribution $N(0, \sigma^2)$.

In the following subsections we present the BER vs. SNR curves obtained from the abovementioned simulations. Because the BER performance of a code on an ISI channel may depend on the transmitted codeword, we determine the generator matrix of the code by means of Gaussian elimination and encode randomly generated sequences of equiprobable bits. For each LDPC code considered, we have included the BER performance obtained from the turbo equalization scheme [15] for comparison purposes. Decoding using turbo equalization was performed in the following way. Each turbo iteration consisted of one pass of the BCJR algorithm followed by $S$ iterations of the sum-product algorithm. If $T$ is the number of turbo iterations, we selected $T$ such that $T(S+1)$ would equal (or be near) the number of iterations $J$ of the PR-BP algorithm. It is worth mentioning that $T$ and $S$ were chosen to achieve the best performance of the turbo equalization algorithm, and were in fact different for different LDPC codes.

### A. Length-495, rate 0.875 (quasi) regular LDPC code

We first consider the MacKay (495,433) code, with rate 0.875, column weight 3 and row weight 24 [26]. This is a slightly irregular code, as it features three parity-check equations with weight 23. This code has been previously considered in the joint-decoding works published in [2] and [19].

Figure 4 depicts the BER versus SNR in decibels. We use the definition of SNR given in (6). In this and in the next examples, a power penalty is applied to the SNR to account for the code redundancy, and is given by $10 \log_{10} R$, where $R$ is the code rate. In this example, the penalty equals 0.58 dB. In Fig. 4, the curve with black square markers shows the BER performance of the code on a memoryless channel obtained making 20 iterations of the sum-product algorithm. This curve may serve as a lower bound to the code performance in a ISI channel. The curve with open circle markers was obtained with the PR-BP algorithm using 40 iterations, while the curve with filled circle markers corresponds to 80 iterations of the same algorithm.

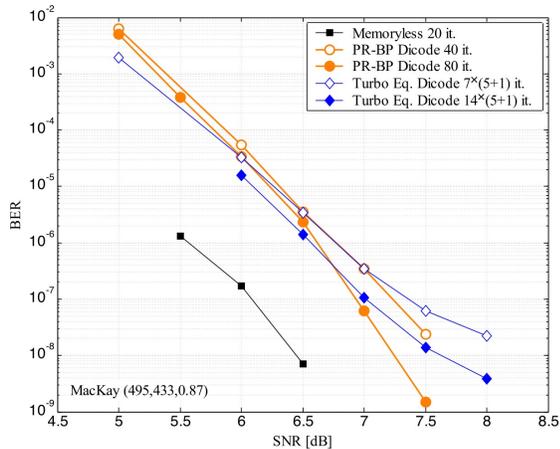

Fig. 4. BER versus SNR in dB of the MacKay(495,433) rate 0.875 (quasi) regular code performing on both a memoryless and a Dicode channel.

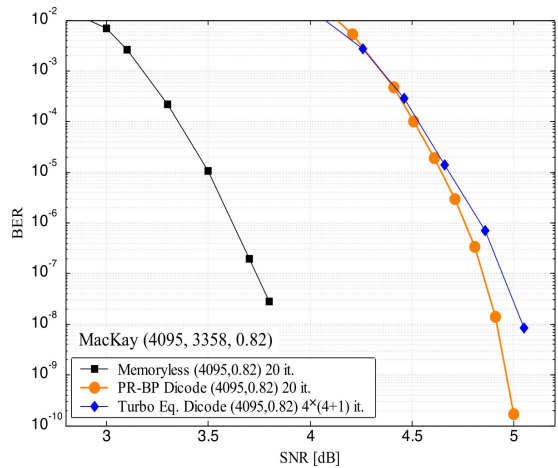

Fig. 5. BER versus SNR in dB of the MacKay (4095,3358) rate 0.82 (quasi) regular code performing on both a memoryless and a Dicode channel.

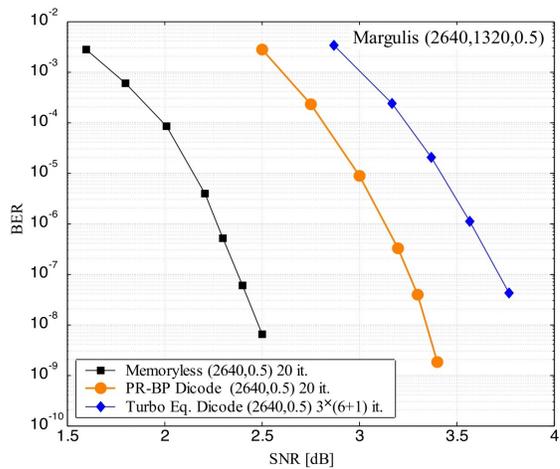

Fig. 6. BER versus SNR in dB of the Margulis (2640,1320) rate 0.50 regular code over both a memoryless and a Dicode channel.

The curves with open and filled diamond markers were obtained using turbo equalization with $8 \times (5+1)$ iterations and $16 \times (5+1)$ iterations, respectively. Note that at high SNR these curves show an error floor, whereas those from the PR-BP algorithm do not. We observe a decoding gain of about 0.5 dB at BER = $10^{-8}$.

### B. Length-4095, rate 0.82, (quasi) regular LDPC code

Our second example considers a code with block length 4095 and rate 0.82 [26]. This code features column weight of 4 and row weights 22 and 23. We simulate this code on both a memoryless channel and a Dicode channel using randomly-generated codewords. Figure 5 shows the BER curves of this code. The memoryless channel is decoded using 20 iterations of the sum-product algorithm, and its BER curve is shown with square markers. The results using the PR-BP algorithm with 20 iterations is depicted by the curve with circle markers. The curve with diamond markers represent the performance of the turbo equalization algorithm using $4 \times (4+1)$ iterations. Note the low BER values achieved with the PR-BP algorithm ($2 \times 10^{-10}$) at SNR = 5 dB. As with the code in the previous subsection, the BER curves on the Dicode channel exhibit a cross-over, with a steeper slope in the case of the PR-BP algorithm. We observe again that with high-rate codes the improvement over turbo equalization can only be seen at low BER. At BER = $10^{-8}$, the decoding gain is 0.15 dB. This gain, however, increases for lower rate codes, as we show next.

### C. Length-2640, rate 0.50, regular LDPC code

The next LDPC code we evaluate is a regular Margulis (2640,1320) code with rate 0.5, column weight 3 and row weight 6 [26]. As with the previous codes we have determined the BER performance on the memoryless and Dicode channels. We use 20 iterations for PR-BP. The best decoding performance using turbo equalization is attained using $3 \times (6+1)$ iterations, different from the optimal combination found for the previous code. We observe that the performance of turbo equalization can be seriously degraded with a careless selection of the iteration parameters $T$ and $S$, as described at the beginning of this Section, at least when the target number of iterations is small.

In the evaluation of this code there is a substantial decoding improvement with the PR-BP algorithm over turbo equalization, as observed in the BER plot of Fig. 6. The BER curves are marked as in the last example. The decoding gain reaches approximately 0.5 dB at BER = $10^{-7}$ and appears to increase at higher SNR. At BER as low as $10^{-9}$ the PR-BP algorithm shows a steadily increasing slope, with no sign of an error floor. We observe a difference of only about 0.8 dB with the memoryless case.

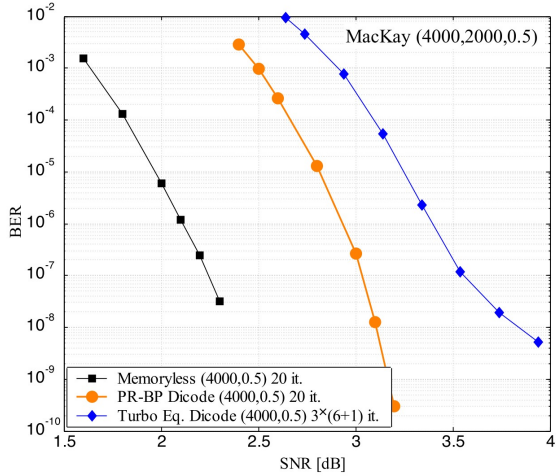

Fig. 7. BER versus SNR in dB of the MacKay (4000,2000) rate 0.50 regular code over both a memoryless and a Dicode channel.

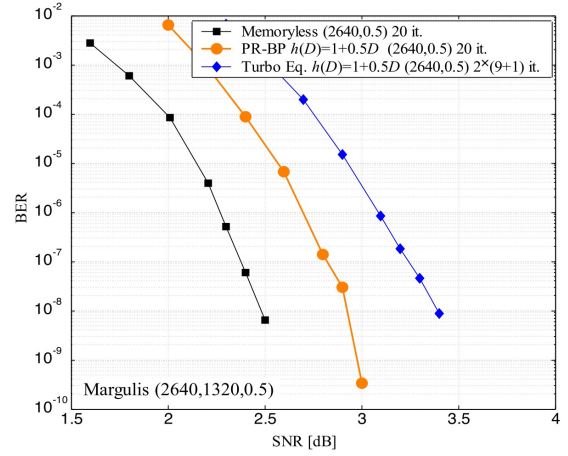

Fig. 8. BER versus SNR in dB of the Margulis (2640,1320) rate 0.50 regular code over both a memoryless and a channel with $h(D) = 1 + 0.5D$.

### D. Length-4000, rate 0.50 regular LDPC code

The last code considered is a MacKay code of block length 4000 and rate 0.5 [26]. This code is regular with column and row weights 3 and 6, respectively. Of the codes considered in this numerical BER evaluation, this is the strongest, as it can be seen in Fig. 7, by the curve with square markers. We have used 20 iterations on the PR-BP algorithm and $3 \times (6+1)$ iterations on the turbo equalization algorithm. The decoding gain achieved by the PR-BP algorithm is about 0.5 dB at BER= $10^{-7}$. It is interesting to see that the turbo equalization algorithm shows an error floor below BER= $10^{-7}$. However, no indication of this is seen with PR-BP, even at BER= $3 \times 10^{-10}$. Because of the error floor, the decoding gain increases to approximately 0.75 dB at BER= $10^{-8}$.

### E. Legth-2640, rate 0.5 LDPC code on PR channel $h(D) = 1 + 0.5D$

As mentioned at the end of Section 6.3, the PR-BP equations are optimal on PR channels with pair-wise ISI. Although not shown here, we have evaluated the Margulis code on a PR2 channel, with $h(D) = 1 - D^2$, and as expected, the BER performance coincides with that of the Dicode channel. It is worthwhile to assess the PR-BP algorithm on a channel with smaller ISI to observe how it approaches to the performance of the sum-product algorithm on a memoryless channel. We have chosen a channel with an arbitrary impulse response $h(D) = 1 + 0.5D$. Figure 8 shows the BER versus SNR of the Margulis code. As one would expect, the BER curve corresponding to PR-BP approaches nicely the curve of the memoryless channel, with a distance of only 0.4 dB. Again, the turbo equalization algorithm required a change of iteration parameters ($T = 2$ and $S = 9$) and its BER performance did not approach that of the memoryless case as fast as the PR-BP scheme.

### F. Complexity and delay

To finish our analysis we briefly present an account of the operations required to complete one iteration of the PR-BP algorithm on a pair-wise ISI channel, of which the Dicode channel is an example. Let $q$ and $p$ be the (constant) out-degrees of the variable nodes and the check nodes of the LDPC code, respectively. In order to simplify the analysis, we do not consider the complexity of the $\tanh$ and $\tanh^{-1}$ functions (as if they were look-up-table operations) and only account for multiplications and additions. We also disregard the computation of the initial likelihoods $u_i$.

To compute $\mu_{i\beta}$ in (33), $(p-2)$ multiplications are required, and only 1 multiplication is needed for $\zeta_{i\aleph}$ in (34). The field $\eta_{i\alpha}$ in (31) requires $(q-2)$ additions of $\mu_{i\beta}$ and 1 addition of $\zeta_{i\aleph}$, besides the 2 explicit additions of the equation. Since there are $(q-1)$ different $\mu_{i\beta}$ and two different $\zeta_{i\aleph}$, the total number of multiplications for $\eta_{I\alpha}$ is $(q-1)(p-2) + 2$. For a pair-wise ISI channel there are only two edges on each ISI node. A careful look at (32) reveals that all the terms in $\eta_{i\aleph}$ have been computed for the field $\eta_{i\alpha}$, or for another field on an edge connected to check node $\alpha$. Only $(q-1)$ additions have to be counted. We also count the 2 additions of the terms in (32). In summary, considering all edges incident on each variable node $i$ (both from the LDPC checks and from ISI nodes), the number of multiplications $N_m$ and additions $N_a$ per symbol per iteration for the PR-BP algorithm on a pair-wise ISI channel are

$$N_m = q(q-1)(p-2) + 2, \quad N_a = q(q+1) + 6. \quad (36)$$

The number of operations per symbol in the BCJR algorithm for a two-state trellis is 18 multiplications and 9 additions. We add to this the cost of the sum-product algorithm, which consists of $q(q-1)(p-2)$ multiplications and $q(q-1)$ additions per symbol. Because the number of sum-product and BCJR iterations within a turbo iteration differs from case to case, the overall complexity varies. Using one of the simulations reported above we compare the complexity and the latency of PR-BP and turbo equalization.

*Example*: In subsection 6.4.4 we simulate the performance of a (4000,2000) LDPC code with weights $q = 3$ and $p = 6$. The total number of operations per symbol performed by the PR-BP algorithm on 20 iterations is 520 multiplications and 360 additions. In the turbo equalization algorithm we have used 3 turbo iterations, each with 7 sum-product iterations and 1 BCJR iteration. This corresponds to 486 multiplications 135 additions per symbol. With respect to delay, the PR-BP algorithm exceeds the performance of the turbo equalization algorithm. To decode a codeword it takes $20\times$ the steps of the PR-BP algorithm, assuming that a parallel architecture is used. In contrast, the same decoding operation takes approximately $3\times N = 12,000$ time steps of the turbo equalization algorithm. Note that the complexity within each time step in the BCJR algorithm is smaller than (18 multiplications and 9 additions) but comparable to that of each symbol in the PR-BP algorithm (26 multiplications and 18 additions).

## V. Conclusions

We have considered the problem of joint channel detection and error correction of LDPC codes over PR channels. We treat the joint PR-LDPC system as an inference problem defined on a graph on which we attempt to determine the marginal probabilities. Finding the marginal probabilities on the combined factor graph is equivalent to decoding a codeword on a PR channel. We have presented a derivation of the belief propagation equations for such a combined system [equations (28),(29)]. The equations originate from the minimization of the Bethe free energy –a well-known technique in statistical mechanics– and provide an optimal solution (limited by the effect of loops in the LDPC part of the graph) for PR channels with polynomial $h(D) = 1 - \alpha D^n$ (where $-1 \leq \alpha \leq 1$) and $n$ is a non-negative integer.

The BP equations for PR channels are explicit and can be solved by any algorithm capable of solving nonlinear equations. We propose a simple iterative algorithm that permits fully parallel implementation on the symbols. Numerical simulations show that the algorithm delivers excellent BER performance on all of the LDPC codes evaluated, surpassing the performance of turbo equalization and showing no error floor above BER= $10^{-9}$. The complexity of the PR-BP scheme in terms of number of operations is comparable to that of turbo equalization. Particularly good characteristics of the PR-BP scheme are its simplicity –as it requires no customization for different LDPC codes– and very low latency. In fact the algorithm exhibits a delay that only depends on the number of iterations, and not on the codeword length. This feature makes it an excellent choice for sequence detection and decoding at high bit rates. Further work is being pursued (a) to determine the optimal BP equations for PR channels with longer memory length, and (b) to analyze, in the spirit of [25], [27], the effects of LDPC-related loops on performance of the PR-BP scheme in the error-floor domain.